\documentclass[twocolumn,showpacs,preprintnumbers,aps,letterpaper]{revtex4}
\input{psfig.sty}
\begin{document}
\addtolength{\topmargin}{1.3cm}
%opening

\title{Thermodynamical properties of dark energy with the equation of state $%
\omega =\omega _{0}+\omega _{1}z$ }
\author{Yongping Zhang}
\affiliation{Department of Physics, Institute of Theoretical Physics, Beijing \ \ \
Normal University, Beijing, 100875, China}
\author{Zelong Yi}
\affiliation{Department of Astronomy, Beijing Normal University,\\
Beijing, 100875, China}
\author{Tong-Jie Zhang}
\affiliation{Department of Astronomy, Beijing Normal University,\\
Beijing, 100875, China}
\affiliation{Kavli Institute for Theoretical Physics China, Institute of
Theoretical Physics, Chinese Academy of Sciences (KITPC/ITP-CAS), P.O.Box
2735, Beijing 100080, People's Republic of China}
\email{tjzhang@bnu.edu.cn}
\author{Wenbiao Liu}
\affiliation{Department of Physics, Institute of Theoretical
Physics, Beijing \ \ \ Normal University, Beijing, 100875, China.}
\email{wbliu@bnu.edu.cn}

\begin{abstract}
The thermodynamical properties of dark energy are usually investigated with
the equation of state $\omega =\omega _{0}+\omega _{1}z$. Recent
observations show that our universe is accelerating, and the apparent
horizon and the event horizon vary with redshift $z$. When definitions of
the temperature and entropy of a black hole are used to the two horizons of
the universe, we examine the thermodynamical properties of the universe
which is enveloped by the apparent horizon and the event horizon
respectively. We show that the first and the second laws of thermodynamics
inside the apparent horizon in any redshift are satisfied, while they are
broken down inside the event horizon in some redshift. Therefore, the
apparent horizon for the universe may be the boundary of thermodynamical
equilibrium for the universe like the event horizon for a black hole.
\end{abstract}

\pacs{98.80.Cq, 98.80.-k}
\maketitle

\section{Introduction}

  Many cosmological observations \cite{1,2,3,4}, such as the type Ia
Supernova(SN Ia), Wilkinson Microwave Anisotropy Probe(WMAP), the
Sloan Digital Sky Survey(SDSS) etc., support that our universe is
undergoing an accelerated expansion. This accelerated expansion is
always attributed to
the dark energy with negative pressure which induces repulsive gravity \cite%
{5,6}. Except for the property of negative pressure, we do not know too much
about the components and other properties of the dark energy. What is more,
we have no idea about how the dark energy evolves. Various dark energy
models have been proposed to describe the evolution, all of which must be
constrained by astronomical observations. In these models, the equation of
state $\omega =p/\rho $ plays a key role and can reveal the nature of dark
energy which accelerates the universe, where $p$ and $\rho $ are the
pressure and energy density of the dark energy respectively. Different
equations of state lead to different dynamical changes and may influence the
evolution of our universe. $\omega$ and its time derivative with respect to
Hubble time $\omega ^{^{\prime }}=d\omega /d(\ln a)$(where $a$ is scale
factor) are currently constrained by the distance measurements of SN Ia, and
the current observational data constrain the range of equation of state as $%
-1.38<\omega<-0.82$ \cite{7}. The cosmological constant model with equation
of state $\omega=-1$, which is considered as the most acceptable candidate
for dark energy has been investigated in \cite{8}. The Quintessence scalar
field with $\omega$ in the range from $-1/3$ to $-1$ has been investigated
in \cite{9,10,11}. The Phantom scalar field with $\omega<-1$, which can
introduce the negative kinetic energy and violate the well known energy
condition, has been found in \cite{12}.

In order to get more information on dark energy, it has been widely
discussed from the thermodynamical viewpoints, such as thermodynamics of
dark energy with constant $\omega $ in the range $-1<\omega <-1/3$ \cite{13}%
, $\omega =-1$ in the de Sitter space-time and anti-de Sitter space-time
\cite{14}, $\omega <-1$ in the Phantom field \cite{15,16} and the
generalized chaplygin gas \cite{17} and so on. More discussions on the
thermodynamics of dark energy can be found in \cite{18,19,20,21}.

In this paper, we investigate the thermodynamical properties of an
accelerated expanding universe driven by dark energy with the equation of
state $\omega =\omega _{0}+\omega _{1}z$ and the Hubble parameter $%
H^{2}(z)=H_{0}^{2}[\Omega _{0m}(1+z)^{3}+(1-\Omega _{0m})(1+z)^{3(1+\omega
_{0}-\omega _{1})}e^{3\omega _{1}z}]$ (where $H_{0}$ is the Hubble constant
and $\Omega_{0m}$ is the parameter of matter density). This model is in
agreement with the observations in low redshift, where the parameter $\omega
_{0}=-1.25\pm 0.09$ and $\omega _{1}=1.97_{-0.01}^{+0.08}$ are suggested by
\cite{22}.

Though combining quantum mechanics with general relativity, it was
discovered that a black hole can emit particles. The temperature of a black
hole is proportional to its surface gravity and the entropy is also
proportional to its surface area \cite{23,24}. The Hawking temperature is
given as $T_{H}=\kappa /2\pi $, where $\kappa$ is the surface gravity of a
black hole, and the entropy of a black hole is $S=A/4$, where $A$ is its
surface area. Hawking temperature, black hole entropy, and the black hole
mass satisfy the first law of thermodynamics $dM=TdS$ \cite{25}. If a
charged black hole is rotating, the thermodynamical first law is expressed
as $dM=TdS+\Omega dJ+V_{0}dQ$, where $\Omega$, $J$, $V_{0}$, $Q$ are the
dragged velocity, angular momentum, electric potential and charge of a black
hole respectively. The event horizon of a stationary black hole is
considered as the system boundary \cite{26}, inside which the black hole
should maintain thermodynamical equilibrium. If our universe can be
considered as a thermodynamical system\cite{27,28,29,30}, the
thermodynamical properties of the black hole can be generalized to
space-time enveloped by the apparent horizon or the event horizon. The
thermodynamical properties of the universe may be similar to those of the
black hole, the thermodynamical laws should be satisfied.

This paper is organized as follows: In Sec II, we examine the
thermodynamical first and the second laws on the apparent horizon with $%
\omega =\omega _{0}+\omega _{1}z$. In Sec III, we similarly examine the
thermodynamical laws on the event horizon. In Sec IV, we draw some
conclusions and discussions.

\section{The thermodynamical laws on the apparent horizon}

The Friedmann-Robertson-Walker metric is \cite{31}
\begin{equation}
g_{ab}=\left(
\begin{array}{cc}
-1 & 0 \\
0 & a^{2}/(1-kr^{2})%
\end{array}%
\right).  \label{y1}
\end{equation}
If setting
\begin{equation}
\tilde{r}=ar,  \label{a4}
\end{equation}%
the dynamical apparent horizon can be determined by the relation%
\begin{equation}
f=g^{ab}\tilde{r}_{,a}\tilde{r}_{,b}=1-(H^{2}+k/a^{2})\tilde{r}^{2}=0,
\label{y2}
\end{equation}
so we can get the apparent horizon radius
\begin{equation}
\tilde{r}_{A}=ar_{A}=1/\sqrt{H^{2}+k/a^{2}}.  \label{xx}
\end{equation}
The WMAP data suggest that our universe is spatially flat with $k =0$, so
the apparent horizon is $\tilde{r}_{A}=1/H$ in this case. The surface
gravity on the apparent horizon is
\begin{equation}
\kappa =-f^{^{\prime }}/2\mid _{r=\tilde{r}_{A}}=1/\tilde{r}_{A},  \label{m4}
\end{equation}%
and the temperature is
\begin{equation}
T_{A}=\kappa /2\pi =1/2\pi \tilde{r}_{A}.  \label{m6}
\end{equation}

In this paper, we concentrate on a more general model with
\begin{equation}
H^{2}(z)=H_{0}^{2}[\Omega _{0m}(1+z)^{3}+(1-\Omega _{0m})(1+z)^{3(1+\omega
_{0}-\omega _{1})}e^{3\omega _{1}z}].  \label{x2}
\end{equation}
As we just want to know more properties about dark energy, we take $\Omega
_{0m}=0$. And the Hubble parameter becomes
\begin{equation}
H^{2}(z)=H_{0}^{2}(z)(1+z)^{3(1+\omega _{0}-\omega _{1})}e^{3\omega _{1}z}.
\label{x3}
\end{equation}
In this section we use temperature, entropy and energy to examine the first
and the second laws of thermodynamics on the apparent horizon. The
temperature and entropy of the universe on the apparent horizon are $%
T_{A}=1/2\pi \tilde{r}_{A}$ and $S=A/4$ respectively.

When calculating the flux of energy through the area $A=4\pi \tilde{r}%
^{2}_{A}=4\pi /H^{2}$ of the apparent horizon, we treat the horizon as a
static surface, but allowing it to vary slowly with time. The amount of
energy flux \cite{15} crossing the apparent horizon within the time interval
$dt $ is \ \

\begin{eqnarray}
-dE_{A} &=&4\pi r_{A}^{2}T_{ab}k^{a}k^{b}dt=4\pi \tilde{r}_{A}^{2}(\rho +P)dt
\nonumber \\
&=&-\frac{3}{2}H_{0}^{-1}\frac{1+\omega _{0}+\omega _{1}z}{1+z}  \nonumber \\
&&\times (1+z)^{-\frac{3}{2}(1+\omega _{0}-\omega _{1})}e^{-\frac{3}{2}%
\omega _{1}z}dz.  \label{c7}
\end{eqnarray}%
The entropy of the apparent horizon is
\begin{equation}
S_{A}=A/4=4\pi \tilde{r}^{2}_{A}/4=\pi \tilde{r}^{2}_{A},  \label{x4}
\end{equation}
and its differential form is
\begin{equation}
dS_{A}=2\pi \tilde{r}_{A}d\tilde{r}_{A}.  \label{a8}
\end{equation}
Thus we can get
\begin{eqnarray}
T_{A}dS_{A} &=&d\tilde{r}_{A}  \nonumber \\
&=&-\frac{3}{2}H_{0}^{-1}\frac{1+\omega _{0}+\omega _{1}z}{1+z}  \nonumber \\
&&\times (1+z)^{-\frac{3}{2}(1+\omega _{0}-\omega _{1})}e^{-\frac{3}{2}%
\omega _{1}z}dz.  \label{c9}
\end{eqnarray}%
From Eq.(9) and Eq.(12), we obtain the result $-dE_{A}=T_{A}dS_{A}$, so the
first law of thermodynamics on the apparent horizon of this
redshift-dependent model is confirmed.

The entropy of the universe inside the apparent horizon is related to its
energy and pressure through
\begin{equation}
TdS_{I}=dE_{I}+PdV.  \label{a10}
\end{equation}%
The energy inside the apparent horizon is $E_{I}=4\pi \tilde{r}^{3}_{A}\rho
/3$ and the volume is $V=4\pi \tilde{r}^{3}_{A}/3$. So we get
\begin{equation}
dS_{I}=\pi \tilde{r}_{A}(1+3\omega )d\tilde{r}_{A},  \label{a11}
\end{equation}%
in which $\omega =\omega _{0}+\omega _{1}z$ and $\tilde{r}_{A}=1/H
=H_{0}^{-1}(1+z)^{-\frac{3}{2}(1+\omega _{0}-\omega _{1})}e^{-\frac{3}{2}%
\omega _{1}z}$.

Thus we can get the derivative of $S_{I} $ with respect to $z$
\begin{eqnarray}
\frac{dS_{I}}{dz} &=&-\frac{3\pi H_{0}^{2}}{2}(1+3\omega _{0}+3\omega _{1}z)
\nonumber \\
&&\times (1+\omega _{0}+\omega _{1}z)  \nonumber \\
&&\times (1+z)^{-3(1+\omega _{0}-\omega _{1})-1}e^{-3\omega _{1}z}.
\label{c12}
\end{eqnarray}%
For the entropy of the apparent horizon, we get%
\begin{equation}
S_{A}=\pi r_{A}^{2}=\pi H_{0}^{-2}(1+z)^{-3(1+\omega _{0}-\omega
_{1})}e^{-3\omega _{1}z},  \label{a13}
\end{equation}%
and the derivation of $S_{A}$ with respect to $z $ is
\begin{equation}
\frac{dS_{A}}{dz} =-3\pi H_{0}^{-2}(1+\omega_{0}+\omega_{1}z)
(1+z)^{-3(1+\omega _{0}-\omega _{1})-1}e^{-3\omega _{1}z}.  \label{a14}
\end{equation}

The total entropy varies with redshift $z $, so its derivation is

\begin{eqnarray}
\frac{d(S_{I}+S_{A})}{dz} &=&-\frac{3\pi H_{0}^{-2}}{2}(1+\omega _{0}+\omega
_{1}z)^{2}  \nonumber \\
&&\times(1+z)^{-3(1+\omega _{0}+\omega _{1}z)-1}  \nonumber \\
&&\times e^{-3\omega _{1}z}.  \label{m3}
\end{eqnarray}
Considering the relation $a=1/(1+z)$, we get
\begin{eqnarray}
\frac{d(S_{I}+S_{A})}{da} &=&\frac{3\pi H_{0}^{-2}}{2}(1+\omega _{0}+\omega
_{1}z)^{2}  \nonumber \\
&&\times (1+z)^{-3(1+\omega _{0}-\omega _{1})+1}  \nonumber \\
&&\times e^{-3\omega_{1}z} \geq0 .  \label{a15}
\end{eqnarray}
We find that the total entropy of the apparent horizon does not decrease
with time and the second thermodynamical law is satisfied on the apparent
horizon.

\section{The thermodynamical laws on the event horizon}

The event horizon of the universe is the position of the greatest distance
that an particle can reach at a particular cosmic epoch, so the definition
of the event horizon is
\begin{eqnarray}
r_{E} &=&a\int_{t}^{\infty }\frac{dt}{a}=\frac{1}{1+z}\int_{z}^{-1}(-\frac{1%
}{H})dz  \nonumber \\
&=&-\frac{1}{1+z}\int_{z}^{-1}H_{0}^{-1}(1+z)^{\frac{-3}{2}(1+\omega
_{0}-\omega _{1})}  \nonumber \\
&&\times e^{-\frac{3}{2}\omega _{1}z}dz.  \label{c6}
\end{eqnarray}

The energy flux through the event horizon can be similarly expressed as

\begin{eqnarray}
-dE_{E} &=&4\pi r_{E}^{2}\rho (1+\omega )dt=-\frac{3}{2}r_{E}^{2}H_{0}(1+%
\omega _{0}+\omega _{1}z)  \nonumber \\
&&\times (1+z)^{\frac{3(1+\omega _{0}+\omega _{1}z)}{2}}e^{\frac{3\omega
_{1}z}{2}}dz.  \label{cc6}
\end{eqnarray}%
Meanwhile, we get
\begin{eqnarray}
dE_{E} &=&\frac{3}{2}r_{E}^{2}H_{0}(1+\omega _{0}+\omega _{1}z)  \nonumber \\
&&\times (1+z)^{\frac{3(1+\omega _{0}+\omega _{1}z)}{2}}e^{\frac{3\omega
_{1}z}{2}}dz.  \label{cc7}
\end{eqnarray}%
The entropy of the event horizon is%
\begin{equation}
dS_{E}=2\pi r_{E}dr_{E}=2\pi r_{E}\frac{dr_{E}}{dz}dz.  \label{b2}
\end{equation}%
Using Hawking temperature of the event horizon, we can also obtain \
\begin{equation}
T_{E}dS_{E}=dr_{E}.  \label{b3}
\end{equation}

\begin{figure}[!t]
\centerline{\psfig{file=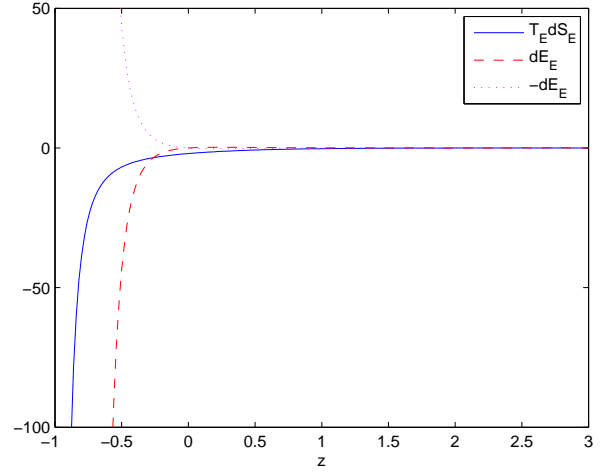,width=3.6in,angle=0}}
\caption{$(T_{E}dS_{E})$, $dE_{E}$ and $(-dE_{E})$ as functions of $z$ with $%
\protect\omega _{0}=-1$ and $\protect\omega _{1}=2$. }
\label{fig1}
\end{figure}

\begin{figure}[!t]
\centerline{\psfig{file=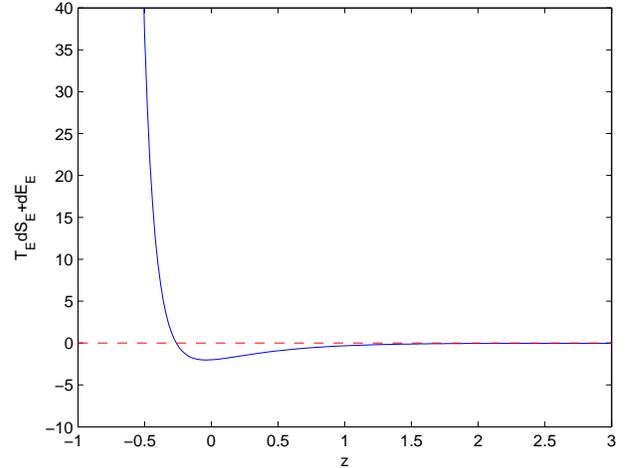,width=3.6in,angle=0}}
\caption{$T_{E}dS_{E}+dE_{E}$ as a function of $z$ with $\protect\omega %
_{0}=-1$ and $\protect\omega _{1}=2$. }
\label{fig2}
\end{figure}

It is shown in Fig.1 and Fig.2 that
\begin{equation}
-dE_{E}\neq T_{E}dS_{E}.  \label{m8}
\end{equation}
In the higher redshift, $-dE_{E}$ is in agreement with $T_{E}dS_{E} $, but
in the lower redshift, they are not consistent. Therefore, the first law of
thermodynamics with the usual definition of entropy and temperature on the
event horizon is not always confirmed.

For the entropy of matter and fluids inside the event horizon, we have%
\begin{equation}
TdS_{I}=dE_{I}+PdV.  \label{b4}
\end{equation}%
The energy inside the event horizon is $E_{I}=4\pi {r}^{3}_{E}\rho /3$, so
we get
\begin{equation}
dS_{I}=2\pi r_{E}^{4}HdH+3\pi r_{E}^{3}H^{2}(1+\omega_{0}+\omega_{1}z)
dr_{E}.  \label{b5}
\end{equation}%
The entropy of the event horizon is $S_{E}=\pi r_{E}^{2}$, so we have
\begin{equation}
dS_{E}=2\pi r_{E}\frac{dr_{E}}{dz}dz.  \label{b6}
\end{equation}%
Similarly we get
\begin{eqnarray}
\frac{d(S_{I}+S_{E})}{dz} &=&2\pi r_{E}^{4}H\frac{dH}{dz}+  \nonumber \\
&&\pi \lbrack 3r_{E}^{3}H^{2}(1+\omega_{0}+\omega_{1} )+2r_{E}]\frac{dr_{E}}{%
dz}.  \label{b7}
\end{eqnarray}%
\

\begin{figure}[!t]
\centerline{\psfig{file=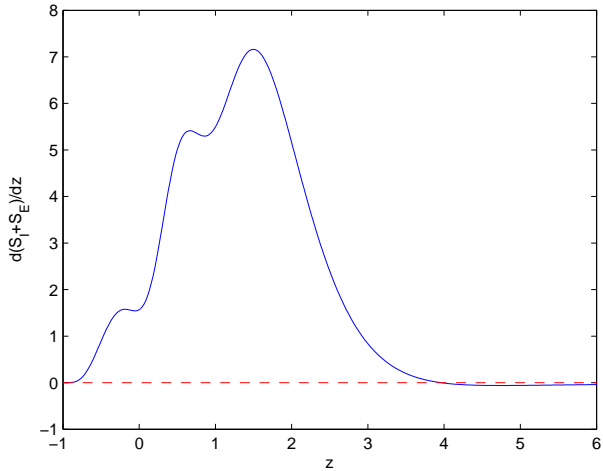,width=3.6in,angle=0}}
\caption{ $\frac{d(S_{I}+S_{E})}{dz}$ as a function of $z $ with $\protect%
\omega _{0}=-1$ and $\protect\omega _{1}=2$.}
\label{fig3}
\end{figure}

In FIG.3, we also find the total entropy inside the event horizon is not
always negative. Through the relation $a=1/(1+z)$, we can get
\begin{equation}
da=-\frac{1}{(1+z)^{2}}dz.  \label{m7}
\end{equation}%
%
%
%
%,
As a result, $d(S_{I}+S_{E})/da$ does not keep positive all the time. The
thermodynamical second law breaks down inside the event horizon.

\begin{figure}[!t]
\centerline{\psfig{file=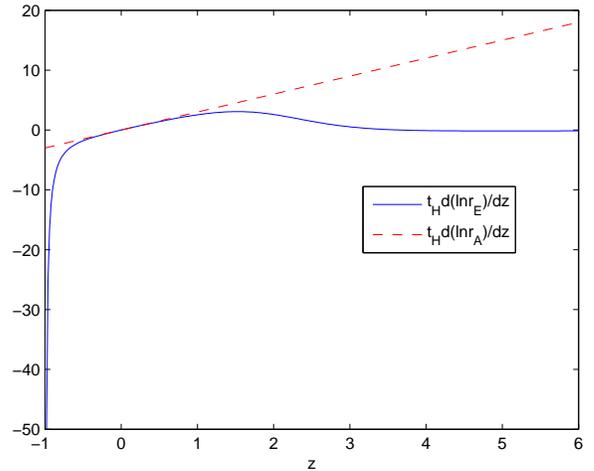,width=3.6in,angle=0}}
\caption{ $t_{H}d(\ln \tilde{r}_{A})/dz$ and $t_{H}d(\ln r_{E})/dz$ with the
redshift during a Hubble scale with $\protect\omega _{0}=-1$ and $\protect%
\omega _{1}=2$.}
\label{fig4}
\end{figure}

\begin{figure}[!t]
\centerline{\psfig{file=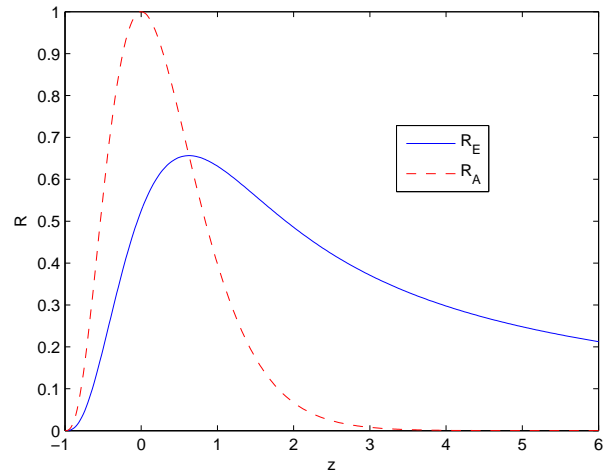,width=3.6in,angle=0}}
\caption{The apparent horizon and event horizon as functions of $z$ with $%
\protect\omega _{0}=-1$ and $\protect\omega _{1}=2$. }
\label{fig5}
\end{figure}

\section{Conclusions and discussions}

\bigskip The Hawking temperature, entropy together with the black hole mass
follow the thermodynamical first law $dM=TdS$, and the Einstein equation can
be derived from the thermodynamical law \cite{30}. As the thermodynamical
laws hold on the event horizon of a stationary black hole, the black hole
keeps thermodynamical equilibrium inside it.

The apparent horizon of a black hole is a two-dimensional surface for which
the outgoing orthogonal null geodesics have zero divergence \cite{32}. A
black hole in asymptotically flat space-time is defined as a region so that
no casual signal (i.e., a signal propagating at a velocity not greater than
light) from light can reach, then the event horizon constitutes the boundary
of a black hole. When the weak energy condition is satisfied, the apparent
horizon either coincides with event horizon or lies inside it. In stationary
black holes, the apparent horizon and the event horizon coincide, such as
Schwarzschild black hole and Kerr-Newman black hole and so on. There always
exists the event horizon located outside the apparent horizon. When a
spherically symmetric black hole becomes non-stationary for some time due to
the matter falling into it, the thermodynamical laws are not fit for a
non-stationary black hole. When a black hole is non-stationary, the apparent
horizon is always not consistent with the event horizon and the
thermodynamical laws may not always hold on the event horizon.

The universe is accelerating and should be not stationary. Due to the heat
through apparent horizon area, we have calculated a variation of the
geometrical entropy on the apparent horizon. We find that the
thermodynamical relation is $-dE=TdS$. In addition, we also examine the
changes of the total entropy with the time $d(S_{I}+S_{A})/da\geq 0$,
therefore, the second law of thermodynamics is evidently confirmed. While
the thermodynamical laws on the event horizon do not hold on, the results
are in agreement with the work of Wang \cite{13}. When the apparent horizon
can be regarded as a thermodynamical system which has associated entropy $%
S=A/4$ with the temperature $T=1/2\pi r$, the differential form of the first
thermodynamical law on the apparent horizon can be written to the
differential form of the Friedmann equation \cite{33}. In a word, the
apparent horizon is a good boundary of keeping thermodynamical properties,
and the universe has a thermal equilibrium inside it. These properties
disappear on the event horizon. Maybe the usual definition of temperature
and entropy are not well-defined , or it is non-equilibrium of
thermodynamics inside the universe.

In FIG.4, in the low redshift, the changing rate of the apparent horizon and
the event horizon is not significant.The thermodynamical description of
horizons \cite{15} will be approximately valid. The apparent horizon always
exists and usually is inside the event horizon. In FIG.5, the event horizon
is outside apparent horizon at higher redshift phase and the universe is
undergoing accelerated expansion. But in this model the redshift becomes
smaller than zero and the equation of state is $\omega <-1,$ the universe
may have a superacceleratd expanding process and the apparent horizon is
outside the event horizon. This model is similar with the Phantom field
which will lead to undesirable future singularity and violate the weak
energy condition.

In summary, it is interesting to investigate the universe from the
thermodynamical viewpoints, although thermodynamical properties need to be
explored more deeply in the future.

\section{Acknowledgments}

Yongping Zhang would like to thank Shiwei Zhou and Xiaokai He for their
valuable suggestions and help. This work is supported by the National
Natural Science Foundation of China(Grant No.10473002), 
the National Basic Research Program of China (Grant No. 2003 CB 716302) and the Scientific Research Foundation
for the Returned Overseas Chinese Scholars, State Education Ministry.

\end{document}